\documentclass[fleqn,usenatbib]{mnras}

\usepackage{newtxtext,newtxmath}
\usepackage[T1]{fontenc}
\usepackage{ae,aecompl}

\usepackage{graphicx}	
\usepackage{amsmath}	
\usepackage{amssymb}	
\usepackage[symbol]{footmisc}
\usepackage{natbib}
\usepackage{cprotect}
\usepackage{wrapfig}
\usepackage{gensymb}
\usepackage{diagbox}
\usepackage{subfig}
\usepackage{stfloats}

\newcommand\miniontot{0.46 }
\newcommand{\minionttoo}{0.37}
\newcommand{\minionottt}{0.06}
\newcommand{\minionotzf}{0.63}
\newcommand{\minionstz}{0.69}
\newcommand{\minionontn}{0.83}

\newcommand\baselinetot{0.23 }
\newcommand{\baselinettoo}{0.16}
\newcommand{\baselineottt}{0.02}
\newcommand{\baselineotzf}{0.4}
\newcommand{\baselinestz}{0.28}
\newcommand{\baselineontn}{0.74}

\newcommand\astrotot{0.69 }
\newcommand{\astrottoo}{0.36}
\newcommand{\astroottt}{0.27}
\newcommand{\astrootzf}{0.90}
\newcommand{\astrostz}{0.90}
\newcommand{\astroontn}{0.81}

\newcommand\pstot{0.70 }
\newcommand{\psttoo}{0.63}
\newcommand{\psottt}{0.30}
\newcommand{\psotzf}{0.92}
\newcommand{\psstz}{0.89}
\newcommand{\psontn}{0.83}

\newcommand\pspop{0.32 }
\newcommand\astropop{0.32 }
\newcommand\minionpop{0.23 }
\newcommand\baselinepop{0.18 }

\newcommand\maga{mag }

\title[Prospecting for X-ray Binary Orbital Periods with LSST]{Prospecting for Periods with LSST -- Low Mass X-ray Binaries as a Test Case}

\author[M. A. C. Johnson et al.]{
Michael A. C. Johnson,$^{1,2}$\thanks{E-mail: Michael.Johnson@soton.ac.uk} 
Poshak Gandhi,$^{1}$ 
Adriane P. Chapman,$^{2}$  
Luc Moreau,$^{3}$ 
\newauthor
 Philip A. Charles$^{1}$\thanks{Leverhulme Emeritus Fellow}, William I. Clarkson$^{4}$ and Adam B. Hill$^{5,1}$
\\ 
$^{1}$School of Physics and Astronomy, University of Southampton, Southampton, Hampshire, SO17 1BJ, UK\\
$^{2}$Electronics and Computer Science, University of Southampton, Southampton, Hampshire, SO17 1BJ, UK\\
$^{3}$Department of Informatics, King's College London, London, WC2B 4BG, UK\\
$^{4}$Department of Natural Sciences, University of Michigan-Dearborn, Dearborn, MI 48128, USA\\
$^{5}$HAL24K Data Intelligence Labs, B3, Johan Huizingalaan 400, 1066 JS, Amsterdam, Netherlands}

\date{Accepted 2018 December 14. Received 2018 December 14; in original form 2018 March 14}

\pubyear{2018}

\begin{document}
\label{firstpage}
\pagerange{\pageref{firstpage}--\pageref{lastpage}}
\maketitle

\begin{abstract}
The Large Synoptic Survey Telescope (LSST) will provide for unbiased sampling of variability properties of objects with $r$\,mag\,$<$\,24. This should allow for those objects whose variations reveal their orbital periods ($P_{orb}$), such as low mass X-ray binaries (LMXBs) and related objects, to be examined in much greater detail and with systematic sampling. However, the baseline LSST observing strategy has temporal sampling that is not optimised for such work in the Galaxy. Here we assess four candidate observing strategies for measurement of $P_{orb}$ in the range 10 minutes to 50 days. We simulate multi-filter quiescent LMXB lightcurves including ellipsoidal modulation and stochastic flaring, and then sample these using LSST's operations simulator (OpSim) over the (mag, $P_{orb}$) parameter space, and over five sightlines sampling a range of possible reddening values. The percentage of simulated parameter space with correctly returned periods ranges from $\sim$\,23\,\%, for the current baseline strategy, to $\sim$\,70\,\% for the two simulated strategies without reduced Galactic sampling. Convolving these results with a $P_{orb}$ distribution, a modelled Galactic spatial distribution and reddening maps, we conservatively estimate that the most recent version of the LSST baseline strategy (\verb|baseline2018a|) will allow $P_{orb}$ determination for $\sim$\,18\,\% of the Milky Way's LMXB population, whereas strategies that do not reduce observations of the Galactic Plane can improve this dramatically to $\sim$\,32\,\%. This increase would allow characterisation of the full binary population by breaking degeneracies between suggested $P_{orb}$ distributions in the literature. Our results can be used in the ongoing assessment of the effectiveness of various potential cadencing strategies.
\end{abstract}




\section{Introduction}

Characterisation of light curve variability is the core of astronomical timing. In particular, there are many classes of unresolved point sources displaying periodic (or quasi-periodic) variations which can be used to reveal information about the source nature. This includes variable, binary and multiple component stars, X-ray binaries and pulsating sources, to name a few.

Low mass X-ray binaries (LMXBs) consist of a neutron star or black hole accreting matter from a low mass (usually < 1\(\textup{M}_\odot\)) orbiting companion donor star via Roche lobe overflow. Approximately 200 LMXBs have been observed within the Milky Way \citep{liu2007catalogue} and $\sim$59 of these are thought to host a black hole, although only $\sim$20 have been dynamically confirmed \citep{corral2016blackcat}. Many of the LMXBs in the \cite{liu2007catalogue} catalogue are either steady or transient LMXBs that have only been seen in outburst. In each case it is the X-ray reprocessed optical emission from the disc which dominates and outshines the donor star. Therefore, compact object characterisation, including orbital period and mass measurement, is most effectively carried out while the LMXB is in quiescence and the donor star can be isolated. 

Quiescent LMXBs are typically far too faint for monitoring with small telescopes, though. This is because they are located throughout the Galaxy with typical distances of order several kpc, or more. Furthermore, high and patchy extinction from gas and dust, especially in the plane of the Milky way, renders them weak and red. So even though studies predict the existence of order $\sim$\,1300 Galactic black hole transients \citep[e.g. ][]{corral2016blackcat}, many are too faint to detect, and the majority remains uncharacterised.

In 2022, the Large Synoptic Survey Telescope (LSST) will begin a 10 year synoptic survey in six filters ($ugrizy$) from Cerro Pach\'on, Chile \citep{ivezic2014lsst}. The high sensitivity and broadband wavelength coverage of LSST will allow it to  probe through Galactic gas and dust in the Milky Way, particularly in the redder filters. Therefore LSST has the potential to expand the known population of LMXB counterparts (down to $r\,\sim\,27$ mag) of which we have only seen a fraction during the short history of X-ray astronomy. The LSST observing strategy will be designed in order to accommodate scientific objectives from a wide range of scientific communities, decisions for this strategy will be supported by the LSST Operations Simulator (OpSim, \citealt{delgado2014lsst}) which generates a complete set of observational metadata for the ten-year simulated mission lifetime. Versions of OpSim are incremented roughly every two years as improvements are made. Whilst LSST has the ability to identify and follow many LMXB counterparts, the current baseline observing strategies (\verb|Minion_1016| for OpSim 3 and \verb|baseline2018a| for OpSim 4)\footnote{\url{https://www.lsst.org/scientists/simulations/opsim/opsim-survey-data} (12/02/18)} include a reduced cadence for fields within the Galactic Plane where LSST is expected to be confusion limited by the high density of sources (defined by \cite{marshall2017science} to be $|b|$ < (1--$l$/90\degree) 10\degree\ for --90\degree <$l$< 90\degree ), the regions in which the vast majority of these systems reside \citep{liu2007catalogue}. 

Other than LSST, the two wide area, broad-band optical survey telescopes that are most suited to observations of LMXBs are the Zwicky Transient Facility \citep{bellm2014zwicky} and Pan-STARRS \citep{2002SPIE.4836..154K}. Although both of these surveys regularly observe large regions of the Galactic Plane, their 5 $\sigma$ single visit depth is several magnitudes brighter than that of LSST and will therefore observe a much smaller fraction of the Galactic LMXB population than LSST. 

An alternative route to LMXB discovery is presented by \cite{casares2018feasibility}, where they present the H $\alpha$ Width Kilo-degree survey (HAWKs) and demonstrate the photometric discovery of LMXBs down to $r$ $\sim$22. Again the optical sensitivity of this survey does not rival that of LSST, however the observations from HAWKs will potentially complement that of LSST for LMXBs by classifying binary systems that are without X-ray follow-up.

At present, very few LMXBs have been observed with sufficient cadence in order to recover $P_{orb}$. One example of the limitations of such a small sample was investigated by \cite{arur2017selection} where they found that the current distribution of LMXB periods could equally be described by two potential period distributions. Furthermore, they deduced that a sample size of $\sim\,275$ LMXB periods would be required to break this degeneracy at the $3\,\sigma$ level.


$P_{orb}$ is the fundamental observable that can be combined with radial velocity information from spectra to determine the masses of the binary components, characterising the masses of LMXBs can gain insight into the processes that form these systems such as the Type Ibc and Type II supernovae. Properties such as the explosion energy, mass cut or the explosion mechanism can all have implications in determining the final mass of the compact object. Additionally, measurements of X-ray binary motion and location in the Galaxy could help characterise the natal kicks that supernovae are expected to impart to the compact object (see: \citealt{van1995galactic}; \citealt{jonker2004distances}; \citealt{repetto2017galactic}; \citealt{gandhi2018gaia}). Knowing $P_{orb}$ is also crucial for ultra-high precision astrometry due to the orbital wobble - when the flux-weighted centroid of emission wobbles at the $P_{orb}$ of the binary system \cite{casares2014j}.


The detection of LMXBs will contribute to two of the four main science drivers of LSST: "exploring the changing sky" and "Milky Way structure and formation", not only are LMXBs transient objects but they can also be used as a proxy to investigate the population of $\sim\,20,000$ black holes that are expected to reside in the Galactic Bulge \citep{miralda2000cluster} as a result of Galactic stellar dynamics.

The results presented in this paper may be useful for the ongoing discussion of optimising the LSST observing strategy. The time of writing also coincides with the call for white papers from the scientific community describing how to strengthen LSST cadence for their scientific requirements. 

By simulating LMXB characterisation in realistic LSST observations, we have investigated the potential for LSST to measure the orbital periods ($P_{orb}$) of these systems in quiescence. We stress that LMXBs are simply a test case of the more generic class of periodic variables which LSST should be able to characterise, so our results can be interpreted more broadly (while keeping peculiarities specific to LMXBs in mind, such as stochastic flaring, described later). 

Section \ref{sec:LMXBLC} outlines the method used for simulating and analysing the data. In Section \ref{sec:results} we discuss the effectiveness of the different observing strategies with regards to period determination and in Section \ref{sec:discussion} we combine these results with distributions describing LMXB periods, magnitudes and Galactic position to find the fraction of the underlying LMXB population for which the LSST can accurately determine periods. 


\section{Methods}
\label{sec:LMXBLC}

In order to examine the ability of LSST to measure the variability properties of LMXBs, first we simulated LMXB lightcurves, and then combined these with simulations of several potential LSST observing cadences to find the expected sampling of the lightcurves. Finally, we used the multi-band Lomb-Scargle algorithm \citep{2015ApJ...812...18V} to recover $P_{orb}$.


\subsection{LMXB Light Curve Simulations}

We simulated quiescent LMXB lightcurves to represent the range of optical counterparts that LSST is expected to observe. $P_{orb}$ and apparent magnitude were varied to encompass a broad area of parameter space outlined by the properties of LMXBs with known counterparts together with the observational constraints of LSST. 

In quiescence, the optical flux of LMXBs is dominated by the companion star, with additional contributions due to the disc and stochastic flaring (see below). The spectral profile was assumed to be of a $K$-type star, a typical late-type companion in many known LMXBs (see e.g. \citealt{casares2014mass}). The spectral profile of a typical $K$-type star\footnote{\url{http://classic.sdss.org/dr5/algorithms/spectemplates/} (ID 11)} was convolved with the LSST's filter transmission coefficients \citep{marshall2017science} in order to calculate the expected magnitudes in the LSST filters. For an object with $r =0$, the full set of LSST magnitudes would be as follows: $u =4.14$, $g =3.24$, $r =0.0$, $i =0.33$, $z =1.05$, $y =2.36$. Note these these magnitudes also include atmospheric transmission effects, which are important at both extremes of the optical spectral regime. In order to account for the additional optical contribution of the disc, which is essentially a flat power-law, a further, constant contribution of 35\% was added to each filter.

To reflect the ellipsoidal modulation expected in an LMXB light curve, a peak to peak brightness variation of 0.1 \maga was assumed. This was split 2:1 between the primary and secondary peaks, chosen so as to be consistent with the sample of quiescent sources published by \cite{zurita2003evidence}. The lightcurves were constructed using alternating portions of two sinusoids with an amplitude ratio of 2:1. The limits of the simulated $P_{orb}$ range were defined to be from 0.0063 days (9 minutes) to 50 days in twenty logarithmically spaced intervals. The minimum value includes the ultra compact LMXBs such as 4U 1820--30 (with $P_{orb}$ = 11 minutes; \citealt{stella1987discovery}) and the maximum includes systems such as GRS 1915+105 (33.5 days; \citealt{greiner2001identification}). The magnitude range used represented the expected, quiescent LMXB magnitude before reddening was applied. This was defined to be from 13 to 22 in the $r$ band. The lower limit to the de-reddened magnitude range corresponds to a typical LMXB with $M_{\rm V}$ = 5 ($M_{r}$ = 4.6) at a distance of 0.48 kpc and the higher limit of 22 corresponds to the same object at a distance of 30 kpc. This range encompasses both the closest candidate LMXB GS\,1354--64 at a possible distance $\sim$\,1\,kpc \citep{gandhi2018gaia} and the farthest edge of the Milky Way from the Sun. 

\begin{figure}
	\includegraphics[width=\columnwidth]{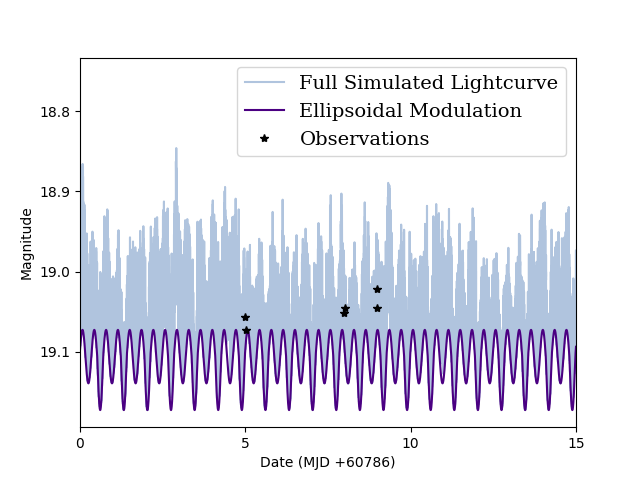}
    \cprotect\caption{Segment of a mock LMXB lightcurve using $r$ band observations of LSST field 1304 with the \verb|astro_lsst_1004_01| observing strategy. The continuous solid purple lightcurve represents the underlying ellipsoidal modulation; light blue includes the additional flaring and noise. Stars symbolise observations made by LSST in the $r$ filter. }
    \label{fig:example_lightcurve}
\end{figure}

LMXBs show additional stochastic flaring behaviour in quiescence, whose origin remains debated. This flaring was simulated according to the flaring power spectra reported by \cite{zurita2003evidence} together with the lightcurve generation algorithm outlined by \cite{timmer1995generating}. The input parameters of the flare algorithm were $\beta$\,=\,--1 representing the slope of the power spectrum and the standard deviation of the flare amplitude was 0.04\,mag. Only flares positive in flux (i.e. brightening the source above the ellipsoidal modulation) are simulated. The simulated amplitude slightly exceeds those exhibited in four of the five systems presented in \citet{zurita2003evidence}, so it is conservative in terms of $P_{orb}$ recovery. The simulated flaring was sampled at 30 second increments as the smallest temporal increment detectable by LSST. The absolute value of the simulated flaring was then taken to reflect the flaring being an additive component to the ellipsoidal modulation. To represent the uncertainties due to shot, instrumental and background noise, we followed the signal to noise ratio prescription suggested by the LSST operations simulation framework\footnote{\url{https://smtn-002.lsst.io/}}. Figure \ref{fig:example_lightcurve} depicts a representative segment from a simulated lightcurve displaying both the final lightcurve as well as the separated contribution from the ellipsoidal modulation alone. 

\begin{figure}
	\includegraphics[width=\columnwidth]{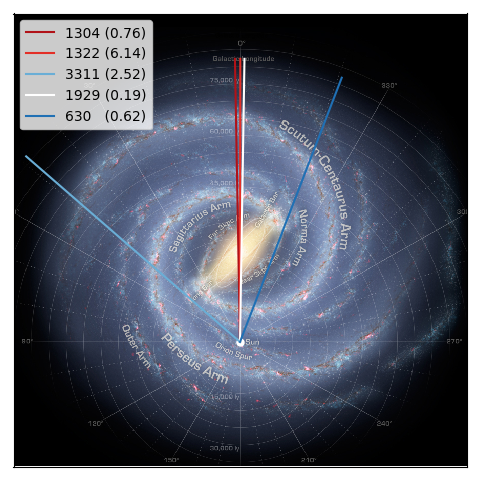}
    \cprotect\caption{Figure depicting the positions of the five chosen LSST fields in the Galactic Plane. The key denotes their LSST field ID and Galactic reddening in \textit{r} magnitudes. (Milky Way image: NASA/JPL-Caltech, ESO, J. Hurt.)}
    \label{fig:fieldpositions}
\end{figure}

Galactic reddening is an important factor to consider, however the clumpy nature of interstellar dust means that this reddening is uncertain, especially within the plane of the Milky Way. As the entire sky could not be simulated within a reasonable time, the reddening to five different LSST fields was used during the simulations. Three fields selected were chosen so as to encompass a wide range of potential Galactic reddenings. A further two were also simulated as they are located such that one field was observed using a different LSST mini-survey and the other resides in the main WFD region. Therefore, they were comprised of fields with OpSim field ID's: 1304, 1322, 630, 1929, 3311. Field 1304 includes the globular cluster NGC 6522 and covers a substantial part of Baade's Window which contains relatively low columns of interstellar dust. Field 1322, corresponds to an LSST field aimed at the Galactic Centre which shows very strong interstellar extinction. Fields 630 and 1929 correspond to two fields which contain famous LMXBs GX 339-4 and Scorpious X-1, respectively. Additionally, field 1929 resides in the main WFD survey region. Finally, field 3311 was included as it is field which resided on the opposite side of the Galactic longitudinal axis to the other chosen fields and well as being located such that it will be observed by the south celestial pole mini-survey. The three Galactic Plane fields were chosen so as to gain meaningful statistics on LSST's $P_{orb}$ recovery in this region and the other two demonstrate how the $P_{orb}$ recovery changes with changing cadence in each observing strategy. The position of these fields in the Galactic Plane is shown in Figure \ref{fig:fieldpositions}. The $E$(B-V) to each target field was found using the dust maps from \cite{schlafly2011measuring} and this was converted to the expected reddening for each LSST filter using the values for $R_{\rm V}$ found also in \cite{schlafly2011measuring}. For each LSST field, the extinction that corresponded to that field was added to the original magnitude range. Observations which had a final magnitude that was either saturating during a single visit or fainter than LSST's 5-$\sigma$ sensitivity limit as described in \cite{marshall2017science} were not used when determining $P_{orb}$. If there were no usable observations in a simulated lightcurve then the period was automatically assumed to not to have been recovered.  

\subsection{Observing Strategy}

\begin{figure}
	\includegraphics[width=\columnwidth]{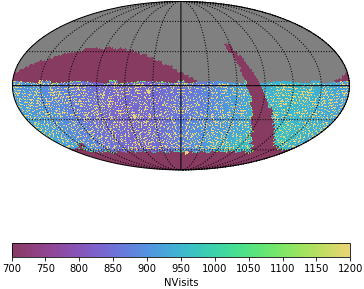}
    \cprotect\caption{ Total number of observations in all bands made using the  \texttt{baseline2018a}  observing strategy, shown in celestial coordinates where zero RA corresponds to the black line in the plane of the y-axis and North=up, East=left. Image credit: http://astro-lsst-01.astro.washington.edu:8080.}
    \label{fig:baseline}
\end{figure}

OpSim \citep{delgado2014lsst} generated mock multi-filter observations \verb|Minion_1016|,  \verb|Minion_1020|, \verb|astro_lsst_01_1004| and \verb|baseline2018a| were downloaded from the LSST simulations page\footnote{\url{http://astro-lsst-01.astro.washington.edu:8081/}, \url{http://astro-lsst-01.astro.washington.edu:8080}}. Figure \ref{fig:baseline} displays all observations made by the new baseline strategy, ,\verb|baseline2018a| (simulated using OpSim 4), of each LSST field over the full ten-year survey, in all filters. In the map, the regions with distinct cadences from the main WFD can be clearly seen in the north, south and Galactic Plane. As with \verb|Minion_1016| (the previous baseline strategy), \verb|baseline2018a| will observe all Galactic Plane fields, in all LSST filters, at a reduced cadence. One key difference between the old and new baseline strategies is that in \verb|Minion_1016|, all Galactic Plane observations occur within the first ten months of operation, whereas these observations are spread out over the ten year survey for \verb|baseline2018a|. \verb|astro_lsst_01_1004| is identical to the baseline strategy \verb|Minion_1016| except that it observes the Galactic Plane with the same cadence as the main survey region. \verb|Minion_1020| utilises a Pan-STARRS-like cadence, with uniform coverage for all observable fields.  Maps showing the total number of observations per field for each observing strategy have been included in Appendix \ref{app:projections}. 

Simulated lightcurves were constructed using the observations that each observing strategy is predicted to make of each target field. The reddening used for each lightcurve corresponded to the line of sight reddening for the field whose observations were used. 


LMXBs enter outburst with recurrence timescales of years to decades, during which the companion star is outshone by the disc and hence the characteristic ellipsoidal modulation cannot be observed. To reflect this in the observations, a randomly selected segment, comprising a consecutive 25\% of the total observing time from the 10 year survey, was removed in all filters for \verb|Minion_1020|, \verb|baseline2018a| and \verb|astro_lsst_01_1004|. However, this was not implemented for \verb|Minion_1016| as all observations of this field occur within the first year. 


\subsection{Multiband Lomb-Scargle Period Measurement}

To take full advantage of the randomly sampled, multi-filter data, the multi-band periodogram outlined by \cite{2015ApJ...812...18V} was used to determine $P_{orb}$. This approach computes the periodogram for each LSST filter separately and regularises them on a common base model to produce a composite. 

The strongest peak in the periodogram is taken to correspond to the orbital period measured for that system and its significance is determined as follows: the dates of all observations and the ellipsoidal modulation magnitudes were shuffled; in order to preserve the red noise inherent in the stochastic flaring, the  flaring magnitudes (in their original order) were then added to the ellipsoidal modulation magnitude; the Lomb-Scargle periodogram was recomputed over this new modified dataset and the power of the maximum peak in this uncorrelated data set was compared to that of the original simulated data. This process was repeated 10,000 times and the significance level was then determined as $\sigma = \frac{x}{N}$ where $x$ represents the number of times that the peak power of the period in the original data was greater than that of the uncorrelated ensemble and $N$ is the total number of shuffles. This formula therefore has a maximum of 1, corresponding to a 100\% recovery rate. If the period was determined incorrectly, defined as $\pm$ 5\% difference between the measured and input periods, the significance was set to zero. This period cut was chosen so as to provide a conservative estimate for $P_{orb}$ recovery herein, and it should be noted that if the period were recovered incorrectly due to aliasing then the correct period may be able to be recovered with further dedicated observations. The decision to keep the flaring magnitudes ordered was motivated as the correlations in the flaring may artificially boost the power of the peaks in the Lomb-Scargle periodogram.      


\subsection{Computation}

All computation was performed on the IRIDIS Compute Cluster nodes at the University of Southampton. The jobs were run on the cluster's nodes which have dual 2.6 GHz Intel Sandybridge processors, 16 CPUs and 64 GB of memory, per node. The total time of computation for all observing strategies was $\sim$16,000 CPU hours.  


\section{Results}
\label{sec:results}
\subsection{Period measurement under the baseline strategy (\texttt{Minion\_1016})}

\begin{figure*}
    \centering 
	\includegraphics[width=\textwidth]{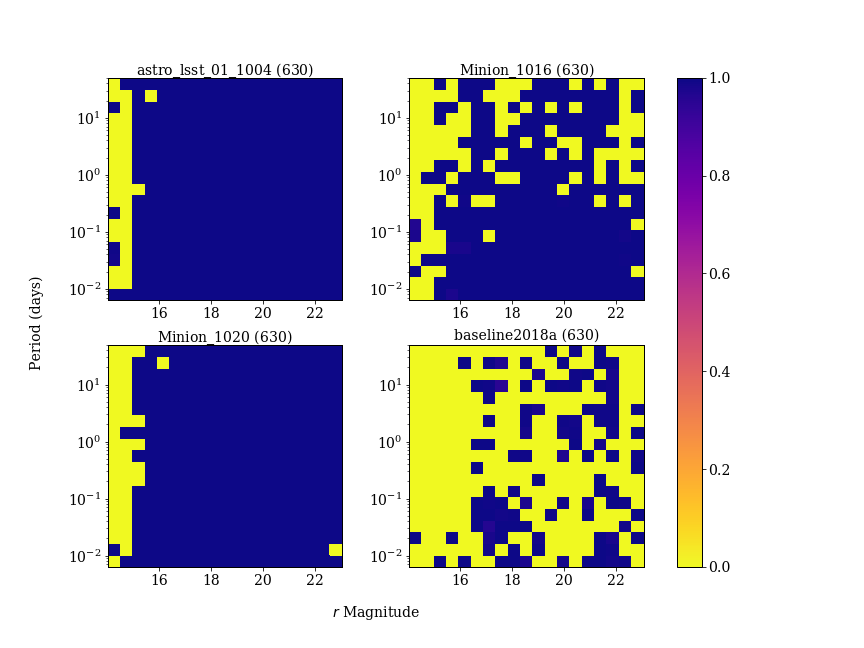}
    \caption{Colour maps displaying the period determination of LMXBs possible in LSST field 630 with observing strategies \texttt{astro\_lsst\_01\_1004}, \texttt{Minion\_1020}, \texttt{baseline2018a} and \texttt{Minion\_1016}. Y axis denotes the orbital period in days, X axis the reddened $r$ mag before adding contributions from ellipsoidal modulation, flaring and noise. The colour denotes the significance of the period detected. If the measured period differed from the actual period by more than 5\%, then the significance was set to zero. The graph shows a bimodality in the significances of period determination as recovered periods that had low significance were often incorrect and manually set to zero. }
    \label{fig:sig}
\end{figure*}

\begin{table*}
	\centering
	\caption {The fraction of the simulated parameter space for which $P_{orb}$ was correctly recovered for each observing strategy, both for the individual LSST fields and the total, combined over all three Galactic Plane fields. The initials denote which cadence was used for that field; South Celestial Pole (SCP), Galactic Plane (GP) or Wide-Fast-Deep (WFD). The reddening is listed in magnitudes. The reddening and coordinates refer to the centre of the field.}
	\label{tab:example_table}
    \setlength\tabcolsep{2pt}
	\begin{tabular}{lcccccr} 
		\hline
      &  & \multicolumn{5}{c}{\underline{~~~~~~~~~~~~Field (Cadence)~~~~~~~~~~~~}} \\
		 & Average & 3311 & 1322 & 1304 & 630& 1929\\ [-2pt]
         &  & (SCP)& (GP) & (GP)  & (GP)  & (WFD)\\
		\hline
{\em Reddening:} & &  &  &  &  &  \\
          $E$(B--V) & & 2.52 & 6.14 & 0.76 & 0.62 & 0.19 \\
          &&&&&&\\
{\em Galactic Coordinates:}  & &  &  &  &  &  \\
$l$\,$(^{\circ})$  & & 49.11 & -0.66 & 0.30 & -21.58 & -1.50 \\
$b$\,$(^{\circ})$  & & 0.80 & -0.90 & -3.49 & -5.24 & 24.81 \\
&&&&&&\\
{\em Observing Cadences: } &&&&&&\\
\texttt{Minion\_1016} & \miniontot & \minionttoo & \minionottt & \minionotzf & \minionstz & \minionontn \\
        \texttt{baseline2018a} & \baselinetot & \baselinettoo  & \baselineottt & \baselineotzf & \baselinestz & \baselineontn\\
		\texttt{Minion\_1020} & \pstot & \psttoo & \psottt & \psotzf & \psstz & \psontn\\
		\texttt{astro\_lsst\_01\_1004} & \astrotot & \astrottoo & \astroottt & \astrootzf & \astrostz & \astroontn\\
		\hline
	\end{tabular}
\end{table*}

In Figure \ref{fig:sig} we show the $P_{orb}$ determination possible with the \verb|astro_lsst_01_1004|, \verb|Minion_1016|, \verb|Minion_1020| and \verb|baseline2018a| observing strategies over the $P_{orb}$-mag parameter space. The simulated observations for this graph were all within the LSST field 630; similar figures covering the other LSST fields have been included in Appendix \ref{app:porbdet}. The magnitude on the x-axis of this figure refers to the mean base $r$ magnitude, as it would be observed after including contributions from reddening for field 630, but without adding any of the introduced stochastic variations. In other words, it corresponds to the mean flux relevant for orbital period determination. The colour denotes the significance of the period measurement and if the period was returned incorrectly, the significance was set to zero. The summaries for the $P_{orb}$ recovery over the full parameter space are shown in Table \ref{tab:example_table}, describing both the prospects per field and averaged over all Galactic Plane fields tested. 

The $P_{orb}$ recovery is worst for the \verb|baseline2018a| observing strategy as it only correctly recovers \baselinetot of the simulated parameter space. This is to be expected as although it has a similar number of observations per field as \verb|Minion_1016|, a 25\% segment of the observations corresponding to potential outburst durations was removed from the full survey lifetime. Therefore, it offered the fewest \textit{usable} observations per Galactic Plane field of any strategy. This is then followed by \verb|Minion_1016| which correctly recovers \miniontot of the parameter space, the low fraction is again due to the relatively small number of Galactic Plane observations per field. The two strategies that performed best were \verb|Minion_1020| and \verb|astro_lsst_01_1004| which correctly recovered $P_{orb}$ for \pstot and \astrotot of the simulated magnitude-$P_{orb}$ parameter space (respectively), averaged over the Galactic Plane fields. The vast majority of the incorrectly recovered periods for both strategies had \textit{no} observations in LSST's visible magnitude range, within that region of parameter space. In these regions, there is no potential for good recovery of $P_{orb}$, regardless of the number of observations.  

\begin{figure}
	\includegraphics[width=\columnwidth]{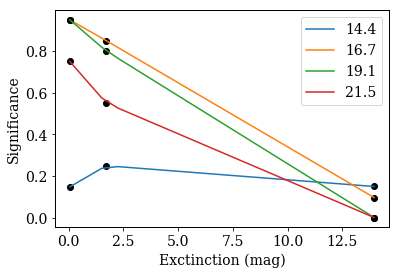}
     \caption{Figure displaying the $P_{orb}$ recovery significance interpolation for the observing strategy \texttt{Minion\_1016} with pre-reddened $r$ magnitudes shown in the key. Each point represents the $P_{orb}$ recovery for a Galactic LSST field, with the significance of recovery on the Y axis and the field's extinction on the X axis. The line represents the corresponding extinction and $P_{orb}$ recovery significance for the twenty chosen, linearly spaced extinction values that are being interpolated.}
\label{fig:redinterpol}
\end{figure}

In order to evaluate the relation between Galactic reddening and period determination, the $P_{orb}$ recovery and reddening were plotted against both magnitude and $P_{orb}$. In order to construct the magnitude-reddening graph, the significance of the $P_{orb}$ recovery was first averaged over all twenty periods for each region of parameter space with a distinct magnitude, field and strategy. This average $P_{orb}$ recovery significance per magnitude was then plotted against Galactic extinction. Each significance-extinction graph therefore had three points, one which corresponded to each Galactic Plane field. These graphs were repeated for all observing strategies. These graphs were then linearly interpolated in order to find the $P_{orb}$ recovery significance at twenty linearly spaced reddening values, ranging from 0 to 13.9 magnitudes. An example is shown in Figure \ref{fig:redinterpol}, each point represents the actual significance of $P_{orb}$ recovery with an LSST field and the line represents the interpolated significance. Using the interpolation, the relation between $r$ magnitude, $r$ band reddening and $P_{orb}$ recovery was then plotted. This process was then repeated, except the average was taken over the magnitudes in order to produce a graph showing the relation between $P_{orb}$, $r$ band reddening and $P_{orb}$ recovery, this figure is included in Appendix \ref{app:redrelation}.


\subsection{Extrapolation to the Underlying Milky Way Population}

In order to calculate the fraction of the underlying LMXB population that LSST is expected to observe with each observing strategy, the above simulations were combined with $P_{orb}$ and magnitude distributions for systems in the Milky Way.

Firstly, to find the expected magnitude distribution, the reddening to each sight-line in the Milky Way was calculated by using the dust map of the Galaxy from \cite{schlegel1998maps}. This reddening was then converted to a \maga ($A_{\rm r}$) assuming an $R_{\rm V}$ of 3.1 and the LSST reddening factors from \cite{schlafly2011measuring}. An absolute $r$ \maga of 4.6 ($M_{\rm V}$\,=\,5) was then used to represent the LMXB quiescent counterpart main sequence $K$-type star. A main sequence $K$-type star was chosen for the companion over a sub-giant because they are typically fainter and will therefore correspond to a more conservative prediction for period determination. Finally, the distribution of systems was assumed to follow the Galactic distribution of LMXBs in the disc and bulge as outlined by Equations 4 and 5 in \cite{grimm2002milky}, combined with the constants from Table 4. The mass ratio used for the disc:bulge was 2:1 and a Milky Way radius of 15 kpc was also assumed. This choice of mass ratio was justified by using the bulge mass estimate from \cite{picaud20043d} and generating a disc mass using Equation 3 and the parameters from Table 2 of \cite{mcmillan2011mass}, giving an approximate ratio of 2:1. The contribution from the spheroid component, as described by Equation 6 in \cite{grimm2002milky} was not included as we were not able to reproduce the mass ratio for it. It is also likely to be a relatively minor contribution to the total mass of BHBs.

The Milky Way was then modelled as a disc with radius 15 kpc (from the Galactic Centre) and height 0.4 kpc, chosen to match the scale height of LMXBs stated in \cite{grimm2002milky}. This disc was then divided into segments using the Galactic coordinate system, \textit{l} and \textit{b} were each segmented in degree intervals and \textit{r} was segmented each 0.1 kpc. The expected probability that an LMXB resided in each section was assigned and these probabilities were integrated over the entire Galaxy and then normalised. The expected reddening and magnitude was then also calculated at each Galactic segment in order to determine what region of the simulated parameter space it corresponded to and therefore, what the significance of $P_{orb}$ recovery in that segment is expected to be. If the region had a magnitude or reddening that was not simulated in the parameter space, then that segment was assigned a $P_{orb}$ recovery significance of zero.    

\begin{figure}
	\includegraphics[width=\columnwidth]{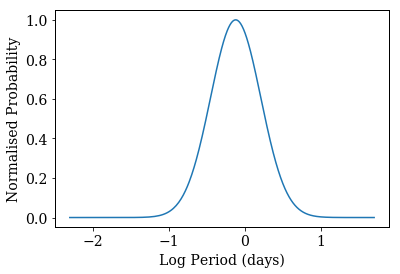}
   \caption{$P_{orb}$ distribution of BHBs, generated by fitting the logarithm of the BHB periods from the BlackCat catalogue \citep{corral2016blackcat}. The probability is normalised to one at peak.}
   \label{fig:periodDistribution}
\end{figure}

The $P_{orb}$ distribution of known systems was characterised by fitting a Gaussian function to the logarithm of the known BHB orbital periods from \cite{corral2016blackcat}. In log space, the distribution had a mean and standard deviation of -0.12 days and 0.47 days, respectively. Figure \ref{fig:periodDistribution} displays the expected BHB $P_{orb}$ distribution calculated using $P_{orb}$ of known BHBs from \cite{corral2016blackcat}. 

The final expected fraction of LMXBs for which LSST could determine periods was then calculated by multiplying the $P_{orb}$ significance, the expected period distribution probability and magnitude distribution probability at all points in parameter space. The equation for this process is shown in Appendix \ref{app:equation}. This was then normalised to a conservative estimate of the LMXB population of 1040 objects to determine the total number of BHB periods that LSST observations could be expected to recover, as shown in Table \ref{tab:pop}. The population estimate used was a combination of the total population estimate of BHBs from  \cite{corral2016blackcat} (1300), combined with the fact that 80\% of known LMXBs reside within the LSST defined Galactic Plane.


\section{Discussion}
\label{sec:discussion}

We have investigated the prospect for periodic signal extraction from LSST light curves. Our test case here is determination of orbital periods of quiescent LMXBs, but our results can be used more generally for assessing various proposed observatory cadencing strategies, especially those relevant for the Galactic plane.

$P_{orb}$ recovery with LSST was shown to be affected by the total number of the observations in the observing strategy; the observing strategies with the highest numbers of observations had the best $P_{orb}$ recovery and those with the fewest had the worst. Observing strategies that did not have a reduced cadence in the Galactic Plane (\verb|astro_lsst_01_1004| and \verb|Minion_1020|) resulted in excellent results for $P_{orb}$ recovery over the simulated mag-$P_{orb}$ parameter space, correctly recovering periods for nearly all of the parameter space which contained observations with magnitudes within the observing range of LSST (within the saturation mag and 5 $\sigma$ limit). Conversely, the baseline strategies (\verb|Minion_1016| and \verb|baseline2018a|) with a reduced cadence in the Galactic Plane, correctly recovered far fewer periods. Furthermore, the current baseline strategy (\verb|baseline2018a|) correctly recovered on average a factor of $\sim 3$ fewer periods than either \verb|astro_lsst_01_1004| or \verb|Minion_1020|. 

The $P_{orb}$ recovery was shown not to vary much between \verb|astro_lsst_01_1004| and \verb|Minion_1020|, even though the former has an additional 100 observations. The most common reason that the period was recovered incorrectly for these two strategies was that there were no usable observations for that region in parameter space (i.e. all observations were so heavily obscured that all observations had magnitudes that were above LSST's 5\,$\sigma$\,limit). In fact, there were very few regions with an incorrectly recovered period where this wasn't the case, meaning that the difference in total observations between these two strategies had little effect. This suggests that the number of observations required for good period determination of LMXBs is higher than the number in either \verb|Minion_1016| or \verb|baseline2018a| however also lower than in \verb|astro_lsst_01_1004| and potentially lower than in \verb|Minion_1020| also. However, as both strategies that had good $P_{orb}$ recovery increase the total number of fields in the Wide-Fast-Deep survey region, without increasing its priority, the median co-added depth achieved is then reduced by 0.04 and 0.15 mags for \verb|astro_lsst_01_1004| and \verb|Minion_1020|, respectively when compared to \verb|Minion_1016| \citep{marshall2017science}. This is a slight reduction in depth whose impact on other scientific programs would need to be assessed.      

One factor that may artificially boost the LMXB recovery for the observing strategy \verb|Minion_1016| is that no observations were removed to account for the potential time that the LMXBs would be in outburst. Removing a randomly selected 2.5 years from this strategy, as was done for the others, is not sensible as the baseline of the observations for this strategy was only 1 year. This will however mean that a fraction of the LMXB population will not be observable through ellipsoidal variability for the entirety of this baseline strategy lifetime, although this is likely to be a negligible effect.

In order to calculate a conservative estimate for the recovered $P_{orb}$ of LMXBs, observations were only considered if they had measured magnitudes within LSST's observable range. However, LSST will perform forced photometry at the location of known objects even if they lie below the $5\,\sigma$ limit during intermediate data releases \cite{juric2018}. This could be relevant for known LMXBs in quiescence that have only been bright enough to be observed in outburst. Therefore, LSST may also be able to determine periods for objects that are outside of this limiting magnitude. There may also be fringe cases where $r$ is $\sim\,24$ mags and its optical variability raises it occasionally above the $5\,\sigma$ limit, thereby producing more usable observations than considered here. However, the combined impact of both of these scenarios is not likely to be significant. 

The overlap and dithering of LSST fields also has the potential to impact the period recovery of LMXBs possible with LSST. These effects could mean that some LMXBs are visible in several LSST fields. Equally, they could also mean that the systems may fall within chip gaps in some images and not others. The impact of these effects will be investigated in the future, however it is not expected to be substantial. Examples of dithering investigations are presented in Chapter 9 (Cosmology) of \cite{marshall2017science}. 

The average and standard deviation value of the flaring had a sizeable effect on the overall period determination. The choice of 0.04 mags for the standard deviation of the flaring was justified as it was representative of the majority of the sample outlined in \cite{zurita2003evidence}. However, one of the sample included a system with standard deviation > 0.1 mags. When implementing the simulations with this value, the significance of $P_{orb}$ recovery decreased significantly. We aim to explore this region of parameter space in the future. 




One should also note that the predictions made by using the dust maps are only estimates as the maps used \citep{schlegel1998maps} represent the integrated reddening along each line of sight, therefore information on the radial change of extinction in the Galaxy is lost. Another limitation to these dust maps is their angular resolution of 6\farcm 1. One should also note that the reddening used per field was used assuming a single pointing, corresponding to the centre of the field, whereas there are potentially many different reddening values per field.


By combining the LMXB period recovery fraction of LSST with a fairly conservative estimate for the LMXB population of 1040 systems, we find a lower limit on the number of systems for which LSST can be expected to determine periods, as shown in Table \ref{tab:pop}. LSST will likely correctly determine $P_{orb}$ for $\sim$200 systems and $\sim$180 systems, while implementing the baseline strategies simulated by OpSim 3 and 4, respectively (\verb|Minion_1016| and \verb|baseline2018a|). Whereas, for observing strategies that do not have a reduced cadence in the Galactic Plane (\verb|astro_lsst_01_1004| and \verb|Minion_1020|), LSST will likely correctly determine $P_{orb}$ for $\sim$300 LMXBs. This sample is sufficient to satisfy the example science case mentioned in the Introduction, as \cite{arur2017selection} deduced that a sample of $\sim$275 LMXB periods would be required in order to distinguish the two different LMXB $P_{orb}$ distributions at the $3\sigma$ level. 

Although LSST will have the {\it capability} to determine the periods for many LMXBs, identification of these sources will require further evidence, and there are numerous other Galactic entities that exhibit similar behaviour to the ellipsoidal modulation of LMXBs. However, there are several potential routes for discerning potential LMXBs. The characteristic X-ray emission seen during the outburst phases of LMXBs can be observed via follow-up with current X-ray telescopes such as Chandra \citep{weisskopf2002overview}. The high sensitivity future instruments such as Lynx \citep{team2018lynx} may also be able to observe the X-ray emission of many of the LMXBs in quiescence. However, X-ray follow-up will not be feasible for all LMXB candidates detected by LSST as X-ray telescopes have relatively small fields of view. All-sky X-ray surveys such as eROSITA \citep{cappelluti2010erosita} will observe the entire Galaxy, however they are limited by their sensitivity and will not be capable of observing the entire Galactic, quiescent LMXB population. 

Another method for identification of large samples has been investigated by \cite{casares2017hibernating}, where they demonstrated that the unique H$\alpha$ emission of LMXBs can be exploited in order to aid their discovery, and they proposed a new survey (the HAlpha-Width Kilo-deg Survey, HAWKs) in order to do so. Furthermore, \cite{casares2018feasibility} investigate the feasibility of HAWKs discerning LMXBs and found that it would be possible for sources down to $r\,\sim\,22$, although, this is $\sim\,2$ magnitudes above of LSST's $5\,\sigma$ depth in the $r$ band.

Spectroscopic follow up can be used for source characterisation and radial velocity determination in order to make mass measurements of the LMXB population. The current generation of spectroscopic telescopes may struggle to observe some of the fainter systems simulated, however this will be feasible with the next generation of instruments available in 2032, after LSST's 10 year lifetime. As the number of LMXBs with dynamically confirmed compact object masses is currently fewer than 20, LSST has the potential to help in improving this by at least a factor of $\sim$10 and potentially a factor of 15+. The implications of this result also extend to many other classes of stellar phenomena involving binary systems which are likely to benefit in exactly the same way as outlined here.   

\begin{table*}
	\centering
	\caption{Fraction of Galactic LMXBs with Measurable Periods as a function of LSST Observing Strategy. The fraction was combined with a total population estimate of 1040 to calculate the total number of systems expected with correctly recovered periods for each observing strategy. }
	\label{tab:pop}
    \centering
    \setlength\tabcolsep{5pt}
	\begin{tabular}{lcccr} 
		\hline
        Observing  & Total Number &Period   &  Period Recovery & Description\\[-3pt]
        Strategy   & of Observations \footnote[1]{footnote} &  Recovery Fraction  &  (No. of systems) &\\
        \hline
        \texttt{Minion\_1016} & 180 &\minionpop & 239 & OpSim 3 baseline\\
        \texttt{baseline2018a} & 134 &\baselinepop & 187 &OpSim 4 baseline\\
        \texttt{Minion\_1020} & 548 &\pspop & 333 & Pan-STARRS-like\\  
        \texttt{astro\_lsst\_01\_1004} & 613 &\astropop & 333 & WFD in Galactic Plane\\
		\hline
	\end{tabular}
    \\Total number of observations represents observations made, averaged over the three Galactic Plane fields \\(minus a 25\% segment for \texttt{Minion\_1016}, \texttt{astro\_lsst\_01\_1004} and \texttt{basline2018a}).\footnote[1]{footnote}
\end{table*}


\section*{Acknowledgements}

We acknowledge the use of the IRIDIS High Performance Computing Facility, and associated support services at the University of Southampton, in the completion of this work. MACJ acknowledges support from ESPRC (EP/N509747/1) and travel support provided by STFC for UK participation in LSST through grant ST/N002512/1. PAC acknowledges financial support from the Leverhulme Trust. PG acknowledges support from STFC (ST/R000506/1). We thank Christian Knigge for discussions on determining the significance of period measurements. We also thank the two referees for their comments and help in improving the quality of the paper.

\vspace{-2em}


\bibliographystyle{mnras}
\bibliography{mnras_template.bib} 




\newpage

\appendix

\section{Observing Strategies}
\label{app:projections}

\begin{figure*}
	\subfloat[]{%
    \begin{minipage}{\textwidth}
	\includegraphics[width=.5\textwidth]{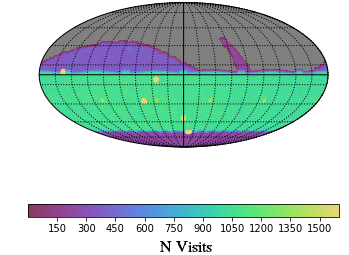}\hfill
	\includegraphics[width=.5\textwidth]{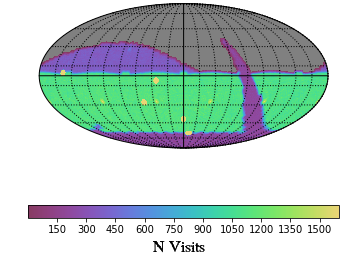}
    \end{minipage}%
	}\par
	\subfloat[]{%
    \begin{minipage}{\textwidth}
	\includegraphics[width=.5\textwidth]{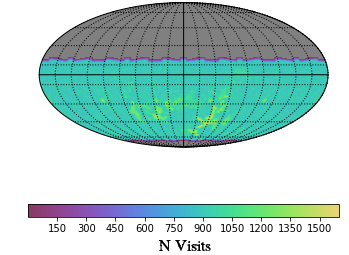}\hfill
	\includegraphics[width=.5\textwidth]{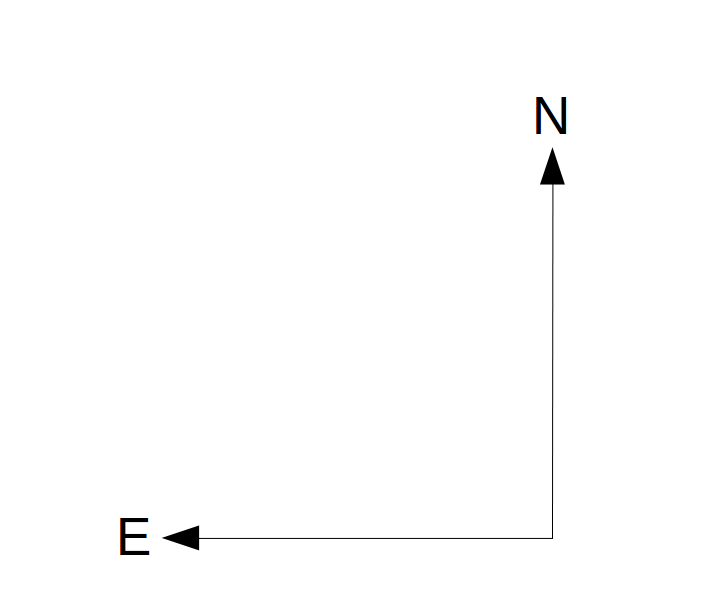}
    \end{minipage}%
    }
     \caption{Total number of observations in all bands made using the \texttt{astro\_lsst\_01\_1004} (left, a), \texttt{Minion\_1016} (right, a), and \texttt{Minion\_1020} (left, b) observing strategies, shown in celestial coordinates where zero RA corresponds to the black line in the plane of the y-axis and North=up, East=left. All graphs were made using the LSST Metrics Analysis Framework.}
\label{fig:observingStrats}
\end{figure*}

Figure \ref{fig:observingStrats} depicts the total number of observations per field made during candidate observing strategies \texttt{Minion\_1016}, \texttt{Minion\_1020} and \texttt{astro\_lsst\_01\_1004} in all bands over the full 10 year survey as simulated with the OpSim \citep{delgado2014lsst}.   

\section{Orbital Period Determination in Simulated LSST Fields}
\label{app:porbdet}

\begin{figure*}
	\subfloat[]{%
    \begin{minipage}{\textwidth}
	\includegraphics[width=.5\textwidth]{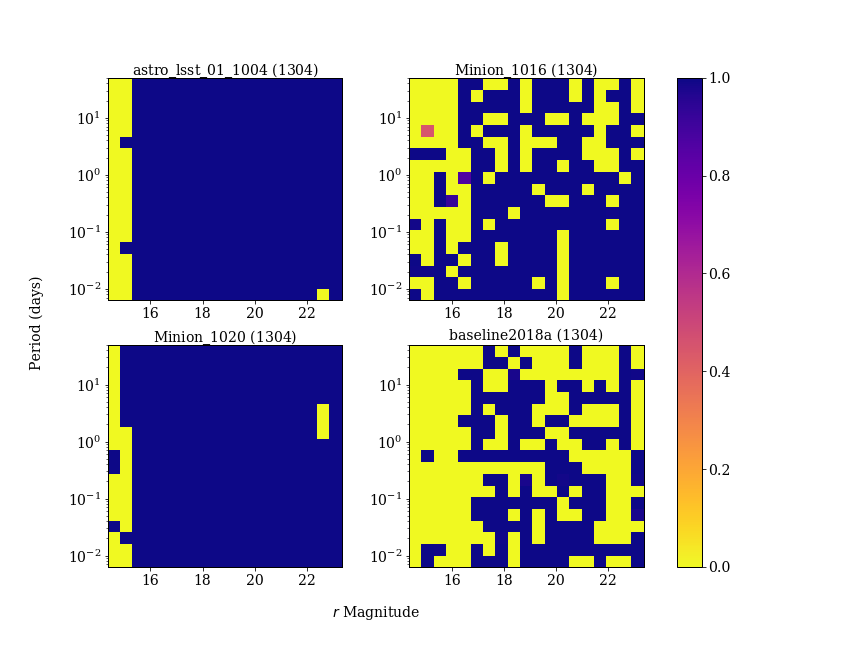}\hfill
	\includegraphics[width=.5\textwidth]{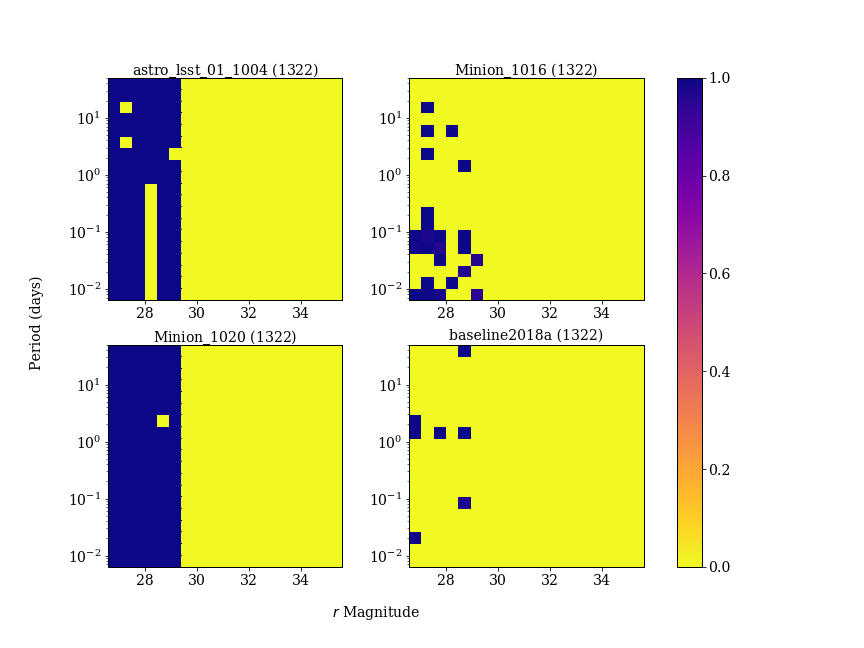}
    \end{minipage}%
	}\par
	\subfloat[]{%
    \begin{minipage}{\textwidth}
	\includegraphics[width=.5\textwidth]{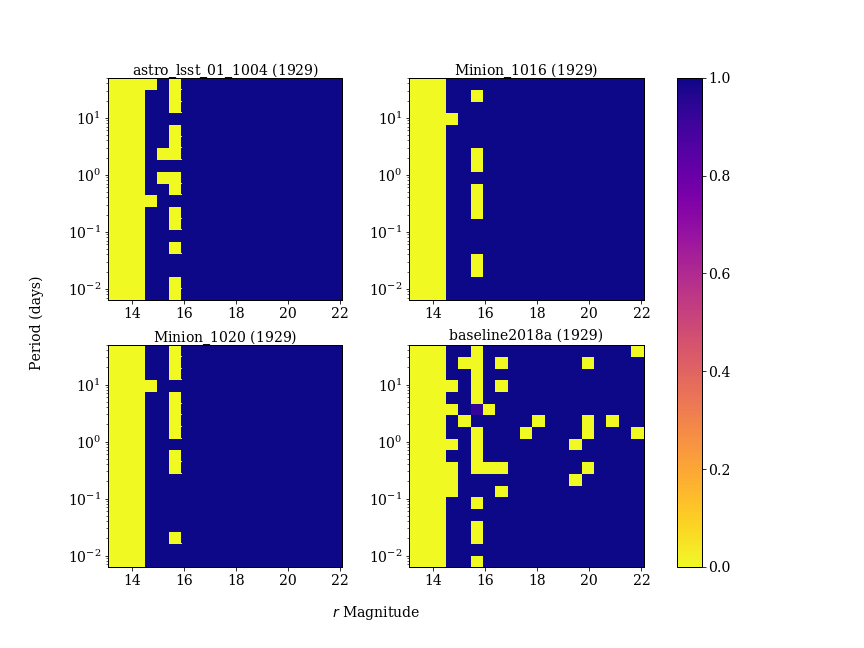}\hfill
	\includegraphics[width=.5\textwidth]{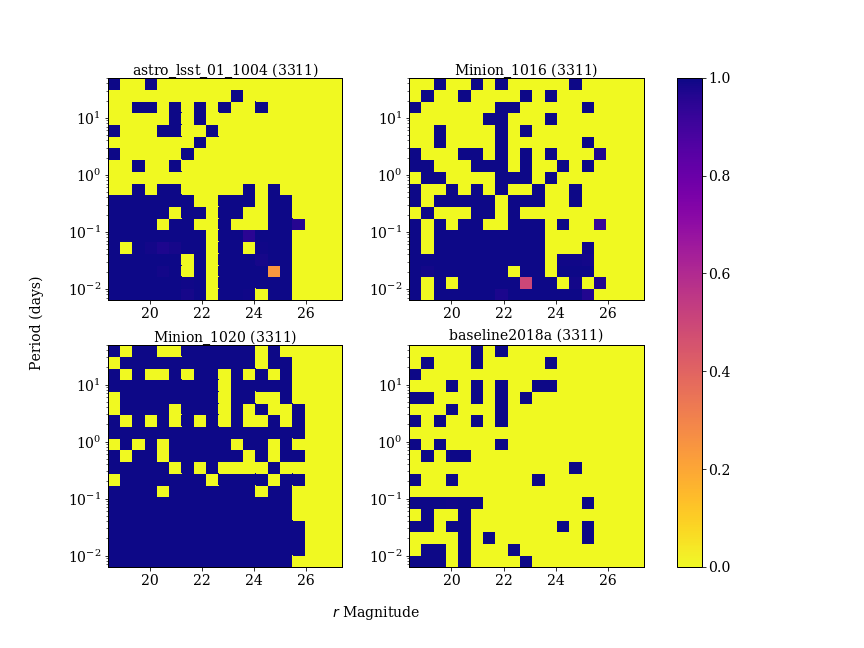}
    \end{minipage}%
    }
     \caption{Colour maps displaying the relationship between magnitude, reddening and period determination of LMXBs possible with observing strategies \texttt{astro\_lsst\_01\_1004}, \texttt{Minion\_1020}, \texttt{baseline2018a} and \texttt{Minion\_1016}. X axis denotes the $r$ band magnitude after reddening had been applied individually for each field and before adding contributions from ellipsoidal modulation, flaring, noise. The X axis denotes the period in days. The colour denotes the significance of the period detected. Simulations using observations of LSST field 1304 (left,a), 1322 (right,a), 1929 (left,b), and 3311 (right,b) are displayed in this figure.}
\label{fig:fullsig}
\end{figure*}

Figure \ref{fig:fullsig} depicts the $P_{orb}$ recovery over the $P_{orb}$-mag parameter space for the four simulated LSST fields which did not have their corresponding diagram included in the main text. The left panel of Figure \ref{fig:fullsig} (b), shows the $P_{orb}$ determination of each strategy with LSST field 1929, which is in the main WFD survey region. This figure demonstrates that when observed with this cadence, the recovery of $P_{orb}$ is very good under all strategies, as there is not the reduced Galactic cadence present. In the right panel of Figure \ref{fig:fullsig} (b), the $P_{orb}$ determination for LSST field 3311 is displayed. This is located such that it will be observed by the South Celestial Pole cadence and the observations are reduced in all strategies except \verb|Minion_1020| due to airmass restrictions. For \verb|Minion_1020|, $P_{orb}$ recovery is reduced only by the relatively high reddening in this field.   

\section{Reddening-Orbital Period and Reddening-Mag Relationships}
\label{app:redrelation}

\begin{figure*}
	\includegraphics[width=\textwidth]{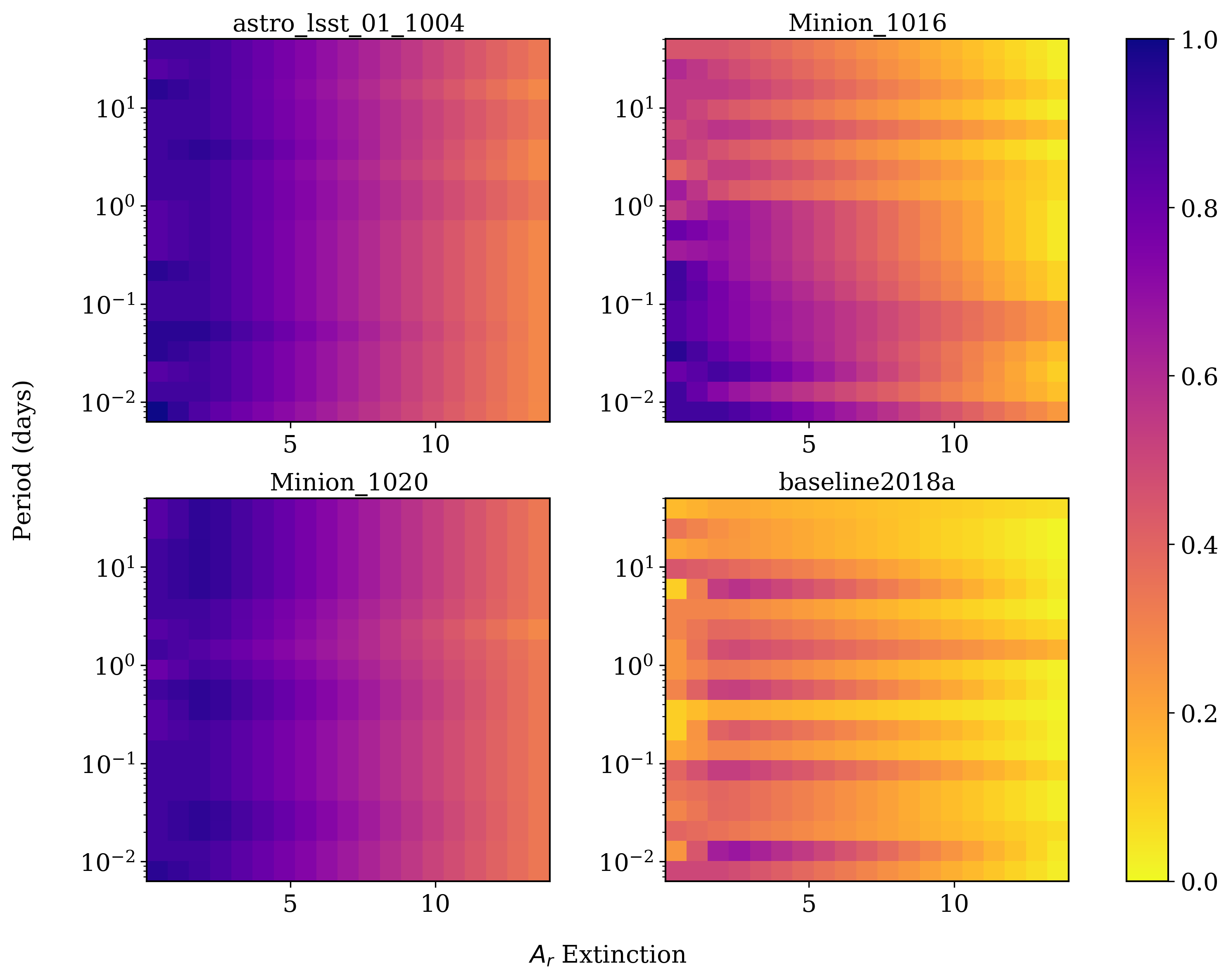}
   \caption{Colour maps displaying the relationship between period, extinction and period determination of LMXBs possible with observing strategies \texttt{astro\_lsst\_01\_1004}, \texttt{Minion\_1020}, \texttt{baseline2018a} and \texttt{Minion\_1016}. Y axis denotes the period in days, X axis $r$ band reddening in magnitudes. The colour denotes the significance of the period detected.}
   \label{fig:allGsRedPeriod}
\end{figure*}

Figure \ref{fig:allGsRedPeriod} depicts the relationship between reddening-period-$P_{orb}$ recovery. We observe a negative correlation between reddening and $P_{orb}$ recovery, which is as to be expected as in most cases, the higher the reddening, the fewer observations in within LSST's visible range are available. The relative lack of bimodality in $P_{orb}$ recovery significance in Figure \ref{fig:allGsRedPeriod} when compared to that in graphs that represent signal fields (Figure \ref{fig:fullsig}) is a result of the extrapolation of $P_{orb}$ recovery between different fields and consequently, the effect of Galactic extinction on $P_{orb}$ recovery.

\section{Galactic Period Recovery Integration}
\label{app:equation}
Equation \ref{equation} details the procedure for summing up the total $P_{orb}$ recovery over the Milky Way. 

$P_{tot}$ is total fraction of the LMXB population that will likely get accurately recovered periods. $r$ represents distance from the Galactic Centre. $P_{p}(P_{orb})$ is the probability of an LMXB having $P_{orb}$, $p$; $P_{m}(M_{obs},r,\theta,\phi)$ is the probability of an LMXB having an observed, post reddening magnitude , $m$ (when calculating $P_{m}(M_{obs},r,\theta,\phi)$, $r$, $\theta$ and $\phi$ were transposed to $l$, $b$ and the radial distance from the Sun using a distance of 7.9 kpc from the Sun to the Galactic Centre). $S_{P_{orb},m}(P_{orb},M_{obs},r,\theta,\phi)$ is the $P_{orb}$ recovery significance with $P_{orb}$, $p$, and magnitude, $m$. $M_{obs}$ is the magnitude of the LMXB before reddening corrections. Finally, $\theta$ and $\phi$ represent angles, in the Galactic Plane and perpendicular to the Galactic Plane, respectively.       

Equation \ref{equation} is held if $12$\,$\leq$\,$M_{obs}$\,$\leq$\,$22$ and $0$\,$\leq$\,$A_{r}$\,$\leq$\,$13.9$. Otherwise, $P_{tot}$\,$=$\,$0$. 
\clearpage
\begin{equation}
\label{equation}
P_{tot} = \int_{0}^{\pi}\,\int_{0}^{2\pi}\,\int_{0}^{15}\,\int_{13}^{22}\,\int_{log(-2.5)}^{log(1.4)}\ r^2 \, sin(\theta)\ P_{P}(P_{orb}) \ P_{m}(M_{obs},r,\theta,\phi) \ S_{P_{orb},m}(P_{orb},M_{obs},r,\theta,\phi) \ dP_{orb}\,dM_{obs}\,dr\,d\theta \,d\phi  
\end{equation}%



\label{lastpage}
\end{document}